 \documentclass[preprint,review,12pt,authoryear]{elsarticle}

\usepackage{graphicx}

\usepackage{amssymb}
 \usepackage{amsthm}
\usepackage{textcomp}  
\usepackage{amsmath}    
\usepackage{multicol}
\usepackage{epsfig}

\biboptions{square}

\journal{Planetary and Space Science}

\begin{document}

\begin{frontmatter}



\title{Proton Cyclotron Waves Upstream from Mars: Observations from Mars Global Surveyor.}


\author[num]{N. Romanelli}
\ead{nromanelli@iafe.uba.ar}
\tnotetext[num]{Corresponding author: Tel.: (5411) 4789-0179 int. 134. Fax: (5411) 4786-8114.}

\author[num]{C. Bertucci}

\author[num]{D. G\'omez}
\author[nus1]{C. Mazelle}
\author[nus]{M. Delva}

 
 \address[num]{Astrophysical Plasmas, IAFE, Buenos Aires, Argentina.}
 \address[nus1]{G\'eophysique Plan\'etaire et Plasmas Spatiaux (GPPS), IRAP, Toulouse, France.} 
\address[nus]{Space Research Institute, Graz, Austria.}


\begin{abstract}

We present a study on the properties of electromagnetic plasma waves in the region upstream of the Martian bow shock, detected by 
the magnetometer and electron reflectometer (MAG / ER) onboard the Mars Global Surveyor (MGS) spacecraft during the period known as Science Phasing Orbits (SPO). 
The frequency of these waves, measured in the MGS reference frame (SC),
is close to the local proton cyclotron frequency. Minimum variance analysis (MVA) 
shows that these 'proton cyclotron frequency' waves (PCWs) are characterized - in the SC frame - by a left-hand, elliptical polarization and propagate almost parallel
 to the background magnetic field. They also have a small degree of compressibility and an amplitude 
{ that decreases with the increase of the interplanetary magnetic field (IMF) cone angle} and radial distance from the planet. 
The latter result supports the idea that the source of these waves is Mars. 
In addition, we find that these waves are not associated with the foreshock and { their
 properties { (ellipticity, degree of polarization, direction of propagation)} do not depend on the IMF cone angle.}
 {Empirical evidence and theoretical approaches 
 suggest that most of these observations correspond to the ion-ion right hand (RH) mode originating
  from the pick-up of ionized exospheric hydrogen.
The left-hand (LH) mode might be present in cases where the IMF cone angle is high.} 
PCWs occur in 62 \% of 
the time during SPO1 subphase, whereas occurrence drops 
to 8\% during SPO2. Also, SPO1 PCWs preserve their characteristics for longer time periods and have greater degree of polarization 
and coherence than those
 in SPO2. We discuss these results in the context of possible changes in the pick-up conditions from SPO1 to SPO2, or steady, spatial
 inhomogeneities in 
the wave distribution. 
The lack of influence from the Solar Wind's convective electric field upon the location of PCWs indicates that, as suggested by
recent theoretical
results, there is no clear relation between the spatial distribution of PCWs and that of pick-up ions.

\end{abstract}

\begin{keyword}
Mars \sep Upstream \sep Cyclotron \sep Waves \sep Pick-up \sep Exosphere

\end{keyword}

\end{frontmatter}

 \def\linenumberfont{\normalfont\small\sffamily}
\section{Introduction}
\label{Introduccion}

Our Solar System is embedded in a plasma flow emanating from the Sun, known 
as the Solar Wind. 
The Solar Wind expands into the Solar System reaching supersonic speeds at a
few solar radii from the Sun. The obstacles in its path can be classified into three types: 
absorbers (for example, unmagnetized asteroids or the Moon) where the Solar Wind interacts directly with their surface, 
 obstacles with intrinsic magnetic fields (such as the Earth and Jupiter), and objects with no
intrinsic magnetic field and  whose atmospheres directly interact with the Solar Wind. In this group we find planets like Venus and Mars, 
Saturn's satellite Titan, and also active comets.

The interaction between the atmosphere of a non-magnetic obstacle and the Solar Wind can be described essentially as a non-collisional interaction 
between a magnetic plasma wind and a 
neutral cloud being ionized by solar radiation and { charge-exchange with the Solar Wind plasma}. An atmospheric obstacle such as Mars's atmosphere 
generates an induced magnetosphere \citep{acuna98,bertucci2005}, 
which has its origin in the exchange of energy and momentum between the Solar Wind and the planetary ions.
The induced magnetosphere is preceded by a bow shock because of the supersonic nature of the Solar Wind. 

However, the interaction can start far beyond the bow shock because particles from the exosphere (mainly hydrogen) \citep{Chaufray} are ionized 
 several planetary radii away from the object.
These particles are ionized mostly by photoionization and charge exchange \citep{modolo05}, which
add a small amount of energy to the ions with respect to their parent neutrals. As the latter are approximately at rest with 
respect to the planet, the ions' planetocentric velocities are also considered to be negligible. 
{ These newborn ions start gyrating around the interplanetary magnetic field (IMF) while preserving 
their parents neutral's parallel velocity. In the planet's frame the gyrating ions also drift perpendicular to $\vec{B}$
and $\vec{E}_{c} (\vec{E}_{c}=-\vec{v} \times \vec{B}).$

The physics of planetary ion pick-up is exactly the same as the one seen at comets and about which there 
is a vast literature [e.g., \cite{mazelleneubabuer93}; \cite{tsu1989}; \cite{tsu1991}; \cite{gary91}, and references therein].
These newborn ions represent a non thermal component of the total ion distribution function which 
is unstable to the generation of electromagnetic waves \citep{wu1972,hartle74}. 
The instability and wave polarization resulting from the pick up process
depends on the IMF cone angle $\alpha_{V,B}$, which is the angle 
between the solar wind velocity ($V_{SW}$) and the IMF at the time 
of pick-up \citep{tsu1986,tsu1987}. If Vsw is parallel to the IMF, the newborn ions will form a beam in the solar 
wind frame and the electromagnetic ion-ion 
right-hand (RH) resonant instability will be predominant \citep{gary93}. On the other hand, if Vsw is perpendicular to the IMF, the newborn ion distribution
 function will drive the  electromagnetic ion-ion left-hand (LH) mode unstable.  
 Both instabilities have maximum growth rates at $\vec{k} \times \vec{B}=0$, where $\vec{k}$ is the propagation wave-number.
At moderate angles of  $\alpha_{V,B}$ the RH instability is still predominant.
\cite{tsubrinc89} found that the maximum growth rate of the LH instability is larger than that of the RH instability for  $\alpha_{V,B}>75$\textdegree, 
whereas, according to \cite{convgary97,garymad}, this cutoff cone angle is  $\alpha_{V,B}=90$\textdegree.

 The ion-ion RH instability  satisfies that the expression
 $ \omega-\vec{k}\cdot\vec{v}^{ion}_{//} +  \Omega_{i}$ is approximately zero for moderate $\alpha_{V,B}$ 
 \citep{gary89,brinca1991}. In this expression $\Omega_{i}$ is the newborn ion gyrofrequency, $\vec{v}^{ion}_{//}$ 
is the ion drift velocity  along the magnetic field $\vec{B}$, and 
 $\omega$ and $\vec{k}$ are respectively the wave frequency and  wave vector in the Solar Wind frame in the case of  propagation parallel to $\vec{B}$.
 As the spacecraft has a negligible planetocentric velocity compared to that of the Solar Wind $\vec{V}_{sw}$, we obtain the 
following expression:
\begin{equation}
 \omega_{sc}=\omega-\vec{k}\cdot\vec{v}^{sc}_{//} \,;\,\,\,\,\,\,\, \vec{v}^{sc}_{//}=-[\vec{V}_{sw}\cdot \hat{k}] \, \hat{k}
\label{e2}
\end{equation}
where $\omega_{sc}$ is the frequency of the plasma wave in the SC 
frame and $\hat{k}=\vec{k}/|\vec{k}|$.

Since the frequency of the electromagnetic plasma wave 
in the newborn ion reference frame is  $\omega_{ion}= \omega-\vec{k}\cdot\vec{v}^{ion}_{//}$, 
and
 the velocity of the planetary particle before ionization is 
negligible with respect to  $\vec{V}_{sw}$, 
we obtain the useful expression:
\begin{equation}
 \omega_{sc}= - \Omega_{i}
\label{eq3}
\end{equation}
where $\Omega_{i} = (q_{i}B/m_{i})$, with $B =$ \textbar$\vec{B}$\textbar $\,$, and $q$ and $m$ being the charge and mass of the newborn ion respectively. 
This means that the waves generated 
by the RH instability are characterized by a frequency which, in the SC frame, is close to
the local ion cyclotron frequency  with a left-handed polarization. 

When the parallel propagating ion-ion LH instability prevails,
 the Doppler correction is very small and therefore the
waves will be observed with a left-hand polarization in both the plasma and SC frames 
at the local ion cyclotron frequency.

}

This suggests that the occurrence of waves at the local ion cyclotron frequency of a particular ion species in the SC frame 
can be associated with the occurrence of the pick up of such ions. 
In this sense, the presence
 of plasma waves is, a priori, an important diagnostic tool for the evidence of ionized exospheric particles.

The first observation of PCWs upstream from Mars' bow shock was made by Phobos-2 \citep{russell90}. These waves
had small amplitudes ($\sim$ 0.15 nT), they were left-hand elliptically polarized in the SC frame, and propagated
at a small angle to the mean magnetic field. PCWs have also been observed 
 by Mars Global Surveyor (MGS) \citep{brain2002,mazelle2004}. The frequency, polarization and propagation angle 
of the waves detected by MGS were similar to those determined from Phobos-2
observations, except their amplitude (2-3 times greater). 
Waves at the local proton cyclotron frequency have also been observed upstream from Venus
 \citep{delva09} and active comets [\cite{tsu1991,mazelleneubabuer93} and references therein].

The eccentric orbits of MGS during the mission's pre-mapping phase \citep{albee2001} allowed observations of PCWs up 
to 15 $R_{M}$ (1$R_{M}$=3390 km: Mars radius). 
\cite{brain2002} performed a statistical analysis of the properties of these waves during the
first aerobraking (AB1) and science phasing orbit (SPO) phases. A few years later, \cite{wr2006} analyzed 85 events
during the AB1 phase and discussed the generation mechanisms at large distances, as well as the possible distribution of waves
depending on the direction of $\vec{E}_{c}=-\vec{v} \times \vec{B}$. Similar analyses were presented by \cite{delva09}, based on measurements provided by 
the magnetometer onboard the Venus Express spacecraft during  
450 orbits.

In this study we carry out an analysis of the PCWs detected by the magnetometer and the electron reflectometer (MAG/ER) onboard MGS
  in the  region upstream from the Martian bow shock during the mission's SPO phase. 
We analyze the frequency, propagation and polarization properties of these waves and discuss the generation 
mechanisms and their relationship to 
the neutral densities at the exosphere of Mars.
We also
analyze the spatial distribution of these waves in a magneto-electric coordinate system centered on Mars (MBE).
Finally, we study the implications of our results and compare them
with recent studies around Mars \citep{wr2006} and Venus \citep{delva2011}.
The article is structured as follows.
In section (2) we describe the capabilities and limitations of MAG/ER in characterizing the properties of the PCW's, 
along with a description of the various methods of analysis that we applied to the measurements. 
{ In section (3) we show typical examples of PCW's as detected by MAG/ER and obtain their properties based on
 the methods described in section (2). 
Next, we show statistical
analyses of the amplitude and the spatial distribution of these waves, as well as the Solar Wind IMF cone angle associated with
them.}
In section (4) we present 
a discussion of the results, and the theoretical approaches that might explain the generation of these waves. 
Finally in section (5) we summarize our conclusions.

\section{Upstream waves: Analysis Methods}
\label{instrumentacion}

\normalfont

MGS entered into 
orbit around Mars on 11 September 1997 \citep{albee2001}. 
During the pre-mapping AB and SPO phases, MGS provided
 measurements of the Martian environment from the unperturbed Solar Wind down to the neutral atmosphere
 from 1683 elliptical orbits. After these orbital phases, MGS reached a final circular mapping orbit at 400 km altitude.

MGS carried a combination of a twin-triaxial fluxgate magnetometer 
system (MAG) and an electron spectrometer used as a reflectometer (ER) \citep{acuna92}. MGS did not carry any instruments dedicated to the measurements of
 ion properties.  The magnetometers (MAG) provided fast measurements (32 vectors/s) over a wide dynamic range (from $\pm$4 nT to $\pm$65536 nT), and the electron spectrometer
 (ER) measured the electron fluxes in 30 logarithmically spaced energy channels ranging from 10 eV to 20 keV with a maximum 
integration resolution of 2 s \citep{er}. In this study we analyze the averaged fluxes over all directions 
with energies higher than 30 eV (below this energy value, the electron distribution function is affected by spacecraft photoelectrons).

The complete removal of spacecraft fields from the upstream MAG measurements was difficult, as the sensors were not mounted on a boom,
but on the outer edge of each of the two solar panels.
However, spacecraft field modeling allowed the reduction of their influence to $\pm 1$ nT \citep{acuna2001}.

We used the MAG/ER measurements from 27 March 1998 through 24 September 1998.
During this period, MGS performed 372 elliptical orbits grouped into two subphases:

\begin{itemize}
 \item From P202 (27/03/98) to P327 (27/05/98). [SPO1 sub-phase]
 \item From P328 (27/05/98) to P573 (24/09/98). [SPO2 sub-phase]
\end{itemize}

The P prefix indicates the periapsis number.

{ The SPO orbits had a constant period of $11.6$ hours with apoapses above the south pole at distances of roughly 6 $R_{M}$. 
They also displayed a monotonic change in their local time from noon to 4AM. During the same period, the Martian season varied from
 early southern hemisphere summer to mid northern hemisphere spring. The SPO1 and SPO2 sub-phases arise as a result of the presence of a 
hiatus due to a solar conjunction around July, 1998. As a result, SPO1 orbits had local times between noon and 10 AM, and SPO2 orbits 
had local times between 10 AM and 4AM. Also, SPO1 orbits are characterized by low to moderate solar zenith 
angles (SZA), whereas upstream 
observations during SPO2 were closer to the terminator plane (see figure 1, \cite{brain2002}).

}

The sampling frequency of MAG measurements analyzed here is $f_{samp}=1.33 Hz$, and due to the influence of spacecraft fields, 
the study is focused on periods where the strength of the IMF
is equal to or larger than 4 nT.

\subsection*{Analysis}  

The analyses performed on MAG/ER data consisted in a characterization of
 their spectral properties, their polarization, their degree of coherence, their
magnetic connection to Mars's bow shock, and their spatial distribution in an electromagnetic coordinate system. A
more detailed description of them is given below.

\subsection{Dynamic spectra and correlation} 

{We generate dynamic Fourier spectra of the magnetic field components in the Mars Solar Orbital (MSO) coordinate system. 
This coordinate system  is centered 
at Mars with its $\vec{X}_{MSO}$ axis pointing toward the Sun, $\vec{Z}_{MSO}$ being 
perpendicular to Mars's orbital plane and positive to the ecliptic north, and while $\vec{Y}_{MSO}$ completing the right-hand system.} 

We also calculate the cross correlation between the fluctuations present in the electron flux measurements for specific energy channels, and 
in the component of the magnetic field along its mean value.

\subsection{Polarization - MVA} 
{ The wave-vector and polarization of PCW's were obtained from minimum variance analysis (MVA)
 \citep{smith76,mvab}}. This method provides an estimate 
of the direction of propagation for a plane wave by calculating the eigenvalues of the covariance matrix 
of the magnetic field within each interval ({ the maximum, intermediate and minimum eigenvalues are $\lambda_{1},\lambda_{2}$ and $\lambda_{3}$, respectively}). 
Then, the wave-vector $\vec{k}$ is associated with the minimum variance
eigenvector. Additionally, the angle between $\vec{k}$ and the mean magnetic field were calculated with an error based on
the number of measurements and eigenvalues ratios \citep{sonerup}.

\subsection{Polarization and coherence - Coherence Matrix} 

An additional tool of analysis consisted in calculating the coherence matrix of the magnetic field for a 0.015 Hz
interval around the local proton cyclotron frequency (MAG uncertainty) \citep{fowler,mcpherron}.
The elements of this matrix provide information about the coherence \citep{tsu2009}, polarization and ellipticity within the selected frequency range. 
These parameters are explicitly indicated in equations (21), (20) and (23), respectively \citep{rk1970}.

\subsection{Magnetic connectivity with the Martian bow shock}
 
We also determine the spatial location of wave observations with respect to the Martian foreshock. For this purpose, we use a static 
model of the bow shock \citep{vignes} and look for an intersection point with the field line
sampled by the spacecraft, assuming a uniform IMF.

\subsection{Spatial distribution} 

Several theoretical studies \citep{modolo05} suggest a dependence of the spatial distribution of pick-up ions upon 
convective electric field  $\vec{E}_{c}$. We investigate if a similar pattern is found on the distribution of waves.
Assuming $\vec{V}_{SW}=-400$ km/s {$\hat{X}_{MSO}$}, we study the spatial distribution of waves (including their amplitude) 
, with respect to 
$\vec{E}_{c}$,
by introducing a \textquotedblleft electromagnetic\textquotedblright coordinate system (MBE) which is centered at Mars and where the 
$\vec{Z}_{MBE}$ axis is parallel to $\vec{E}_{c}$, $\vec{X}_{MBE}$ is antiparallel to $\vec{V}_{SW}$, and $\vec{Y}_{MBE}$ completes the right hand triad.
The IMF cone angle is also included in this analysis.

These methods have been applied on selected time intervals to illustrate the main properties of these waves and then
on 10-minute overlapping intervals in order to study the changes in the waves' properties along the MGS trajectory.
The duration of these segments was chosen so as to contain a significant number of proton cyclotron periods.

\section{Results}
\label{resultados}

\begin{figure}[!ht]
\begin{center}
\includegraphics[scale=.45]{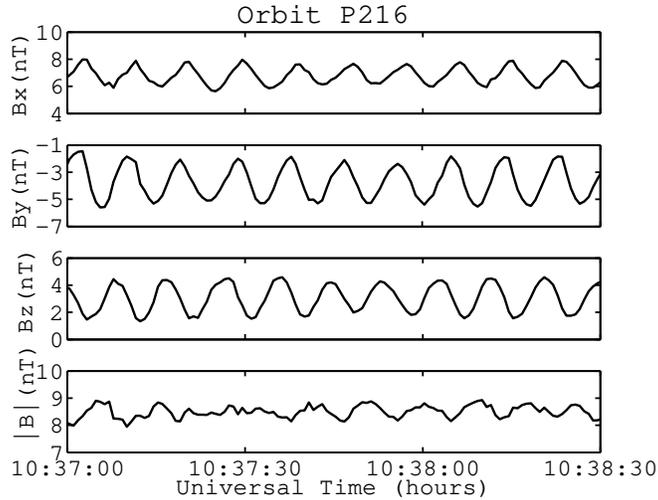}
\caption{Magnetic field measurements during part of the orbit P216 in MSO coordinates, (April 3, 1998).}
 \label{medicionesb}
\end{center}
\end{figure}

\begin{figure}[ht]
\begin{center}
\includegraphics[scale=.45]{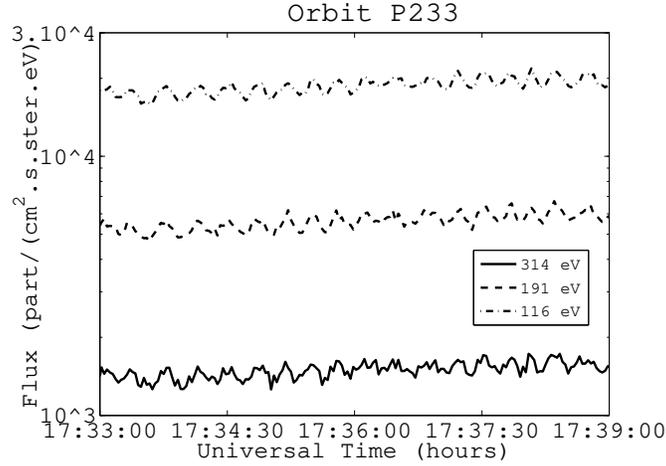}
\caption{Electron flux measurements during part of the orbit P233, (April 11, 1998).}
 \label{medicionesflux}
\end{center}
\end{figure}

 Figures \ref{medicionesb} and \ref{medicionesflux} show typical examples of wave activity present in MAG/ER measurements.
Figure \ref{medicionesb} illustrates an example of PCW's seen by MAG upstream from the Martian bow shock. The top three panels show the magnetic field components in the MSO reference frame. 
At first glance, the oscillations show a well-defined frequency, large amplitudes (up to 4 nT) and a high degree of
coherence. 
 In this interval, the { average} magnetic field ($\vec{B}_{o}$) is quite steady: its magnitude is slightly above 8 nT and
 its orientation makes an angle of 34\textdegree with respect to the Mars\texttwelveudash Sun direction. 
Figure \ref{medicionesflux} shows the electron fluxes measurements seen by ER for energies 116, 191 and 314 eV (with maximum time resolution of 2s). 
Oscillations with features similar to those measured by MAG can be observed, { which 
 we assume also corresponds to fluctuations in the total electron density.}

\begin{figure*}[ht]
\begin{center}
\includegraphics[scale=.4]{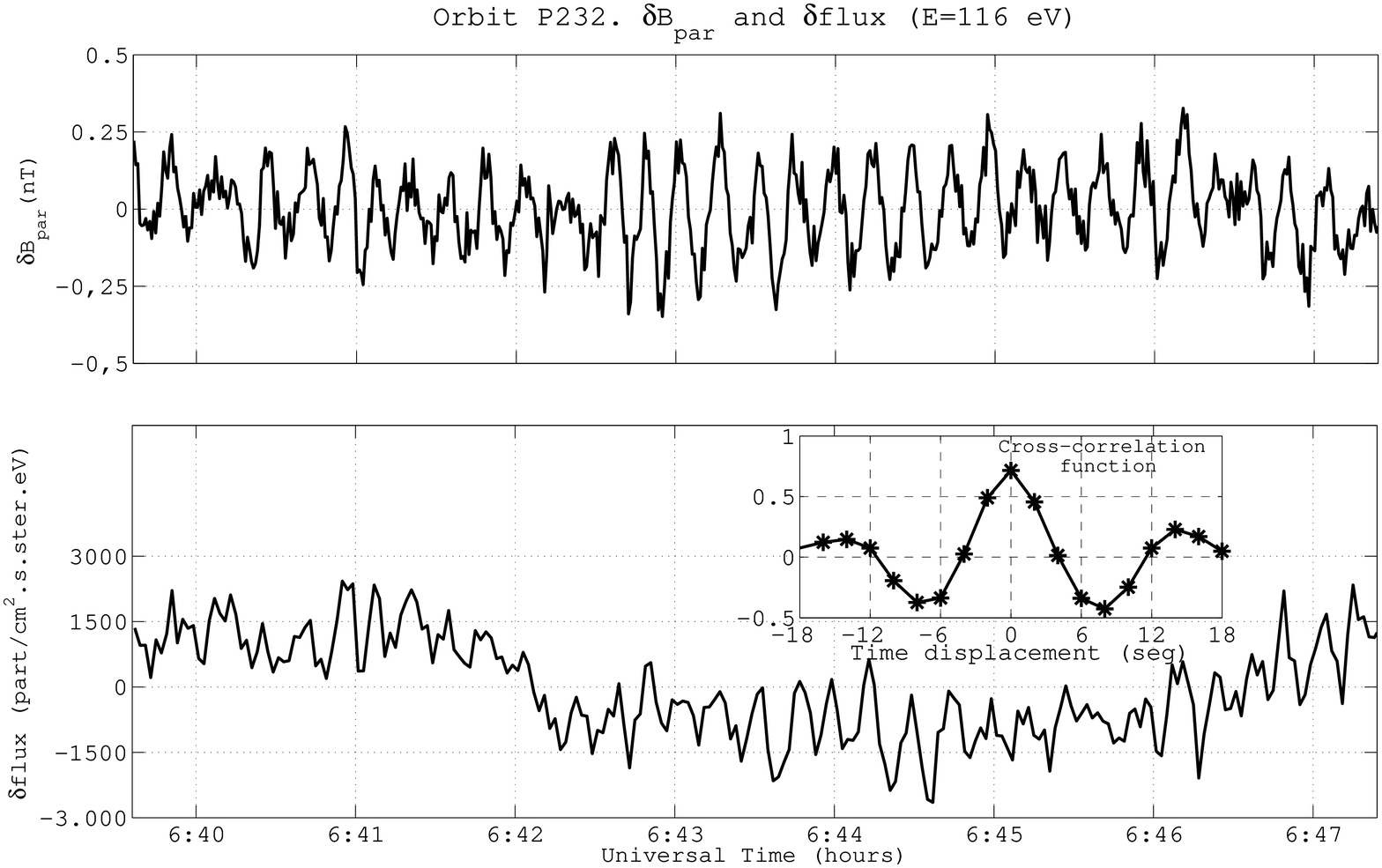}
\caption{Fluctuations in the parallel component of the magnetic field and in the electron flux measurements (E=116 eV) during part of the orbit P232, (April 11, 1998).}
 \label{crosscor}
\end{center}
\end{figure*}

\subsection{Temporal spectra} 
When simultaneous MAG and ER measurements are compared, we find that the component of the magnetic field parallel 
to $\vec{B}_{o}$ and the omnidirectional electron flux are in phase. 
Figure \ref{crosscor} shows the fluctuations in the parallel component of the field (upper panel), and in
the electron flux at 116 eV (lower panel) during part of the orbit P232.
The cross-correlation of the two time series shown in the enclosed panel 
 displays a peak at zero displacement with a correlation value of 0.7,
falling down to 0.44 at $\pm2$s. 
We repeat this analysis for other events and for different energy channels 
with similar results, which is consistent with both time series being in phase, considering an error of
$\pm2$s associated with the cadence of the ER instrument.

\begin{figure}[ht!]
\begin{center}
\includegraphics[scale=.27]{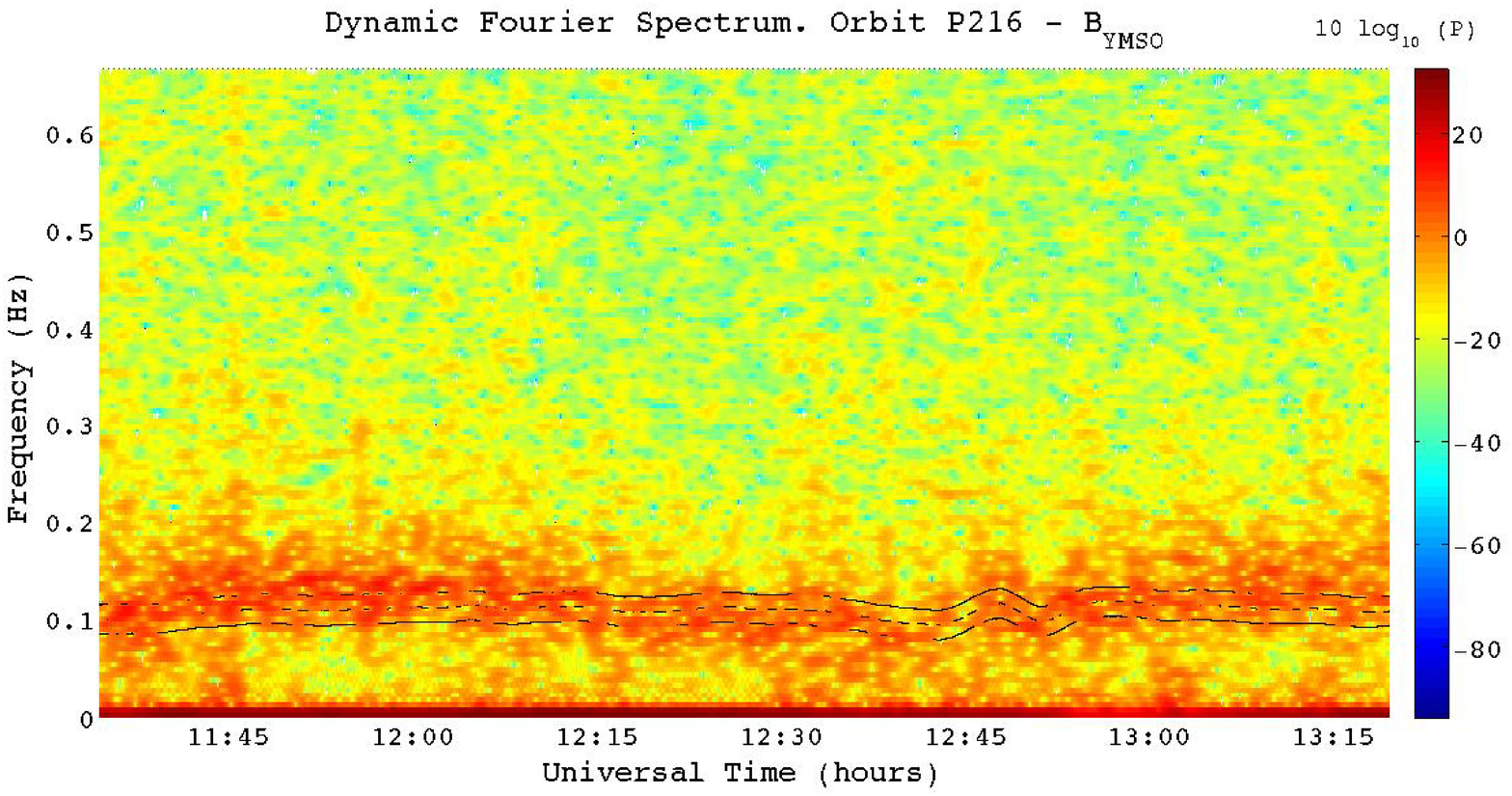}
\caption{ Fourier dynamic spectrum of ${B}_{YMSO}$, Orbit P216, (April 3, 1998).
The calculated local proton cyclotron frequency from MAG data is plotted in black for reference (line in the middle). The 
black lines at the sides correspond to the error bars associated with the MAG data. P is the power spectral density.}
 \label{espectrograma}
\end{center}
\end{figure}

We also find that most of the waves observed by MGS MAG/ ER have frequencies 
in the SC frame which coincide with the local proton cyclotron frequency ($\Omega_{p}=B/m_{p}$), within a $\pm 1$ nT uncertainty. 
Figure \ref{espectrograma} shows a dynamic Fourier spectrum of the $\vec{Y}_{MSO}$ component of the magnetic field (${B}_{YMSO}$) in the orbit P216.
The peak in the frequency of the oscillations is systematically close to the calculated proton cyclotron
frequency for several hours.

\subsection{Polarization - MVA} 
\begin{figure}[ht]
\begin{center}
\includegraphics[scale=0.5]{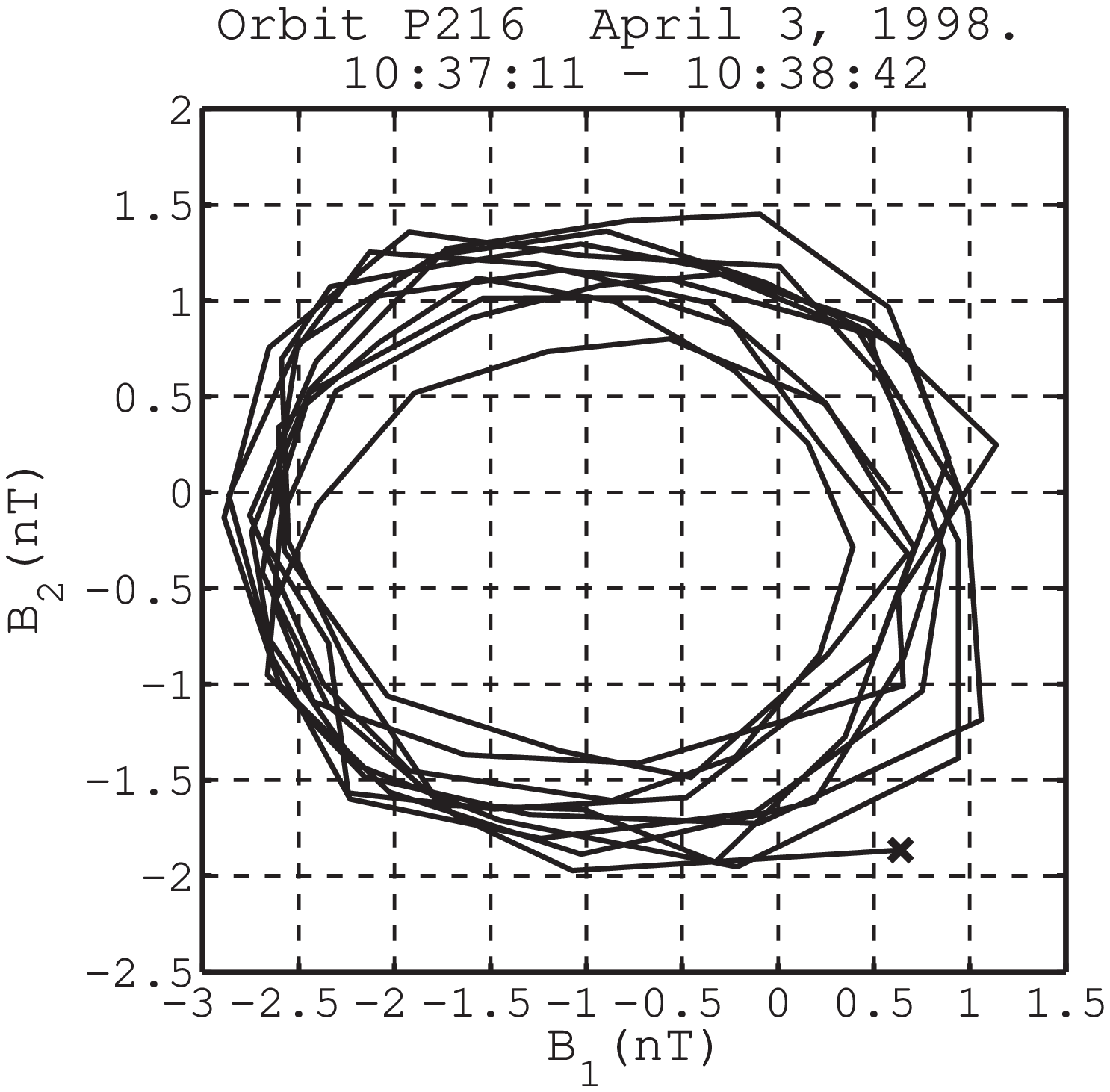} 
\caption{MVA hodogram corresponding to a 10 cyclotron period interval during orbit P216. 
 The mean magnetic field points out of the page ($\vec{B}_{o}= [-1.02, -0.18, 8.27]$ nT). The cross indicates the start of the time series analyzed.}
 \label{mvab}
\end{center}
\end{figure}

MVA shows that the { waves are planar, propagate
 almost parallel to the background magnetic field and have a left-hand circular polarization}
 (in the SC frame) with respect to $\vec{B}_{o}$. 
Figure \ref{mvab} shows the results of MVA applied on 10 cyclotron periods of data during orbit P216. 
In this case, MVA yields a high $\lambda_{2}/\lambda_{3}$ ratio (50.3) and a { small angle between $\vec{k}$
and $\vec{B}_{o}$ ($\theta_{kB} = 8$\textdegree).}
The hodogram describing the maximum and intermediate variance components clearly shows a left hand polarization, as 
the mean magnetic field for this interval points out of the page.

In the case of orbit P216, the propagation properties of the waves remain unchanged for several hours.
As a result, an estimation of the $\theta_{kB}$ from 10 proton cyclotron periods windows for which
 $\lambda_{2}/\lambda_{3} > $ 10 \citep{knetter04} yields a value of 9.9\textdegree.

We also find that these waves 
typically have more power perpendicular to the mean-field direction than along it, which again indicates that $\vec{k}$ makes a
 small or moderate angle with the mean-field direction.

\subsection{Polarization and coherence - Coherence Matrix} 

\begin{figure}[ht]
\begin{center}
\includegraphics[scale=0.47]{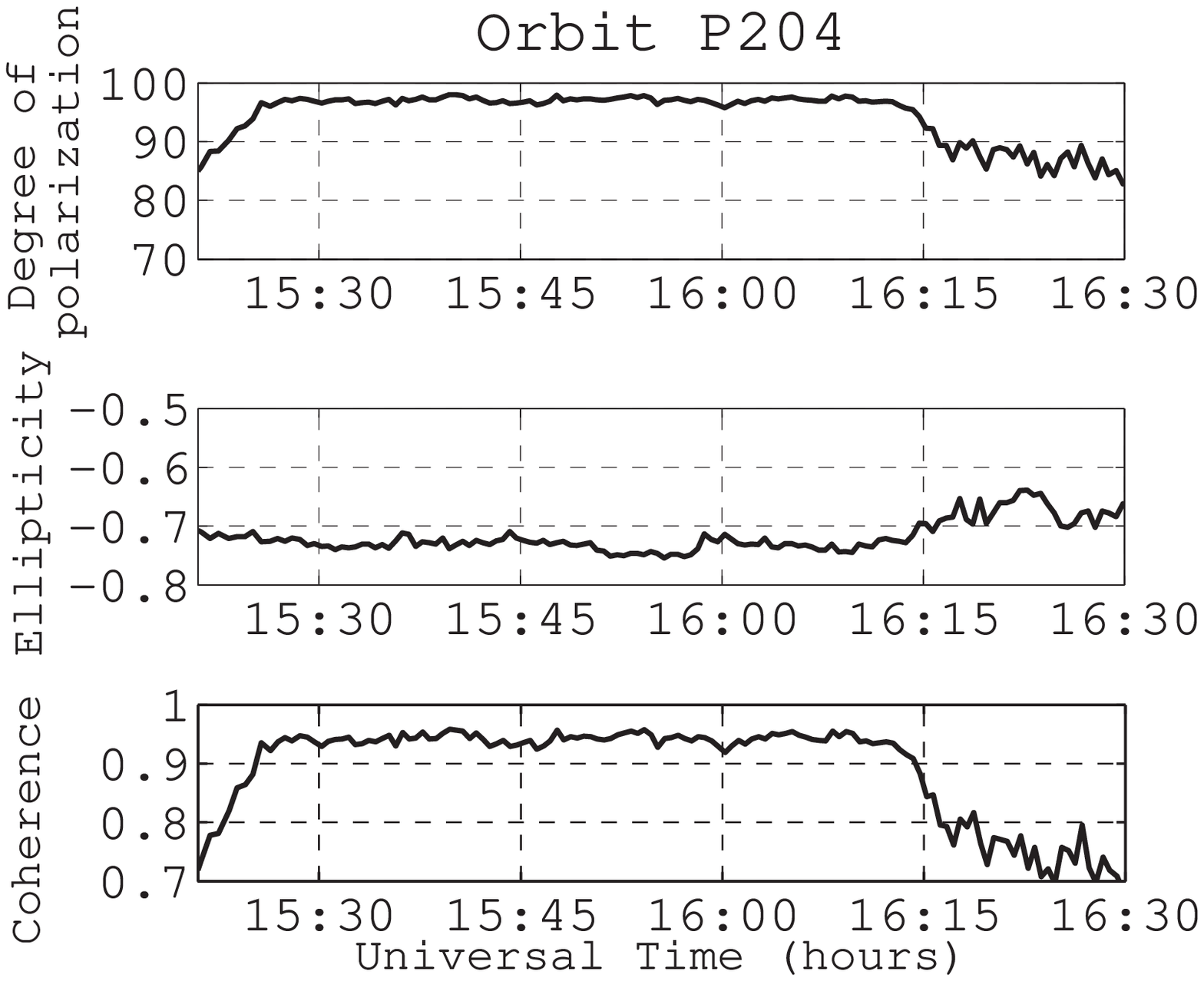} 
\caption{  Properties of the ultra low frequency waves observed by the MGS spacecraft 
during
a part of the orbit P204, (March 28, 1998).}
 \label{otrasprops}
\end{center}
\end{figure}

The apparent coherence of the waves shown in Figure \ref{medicionesb} is confirmed with an estimate of the coherence coefficient described in section (2).
 Figure \ref{otrasprops} illustrates an example of PCW's (orbit P204) with 
 a high degree of coherence ($\geq 0.7 \,$) and polarization ($\geq 70\%$) for over an hour. 
Their ellipticity coefficient indicates
a left-hand elliptical  polarization ($<$ -0.6).
Similar analyses applied to other SPO1 orbits suggest that these properties also remain unchanged for long periods of time (typically 1 hour).

\begin{figure}[ht]
\begin{center}
\includegraphics[scale=0.5]{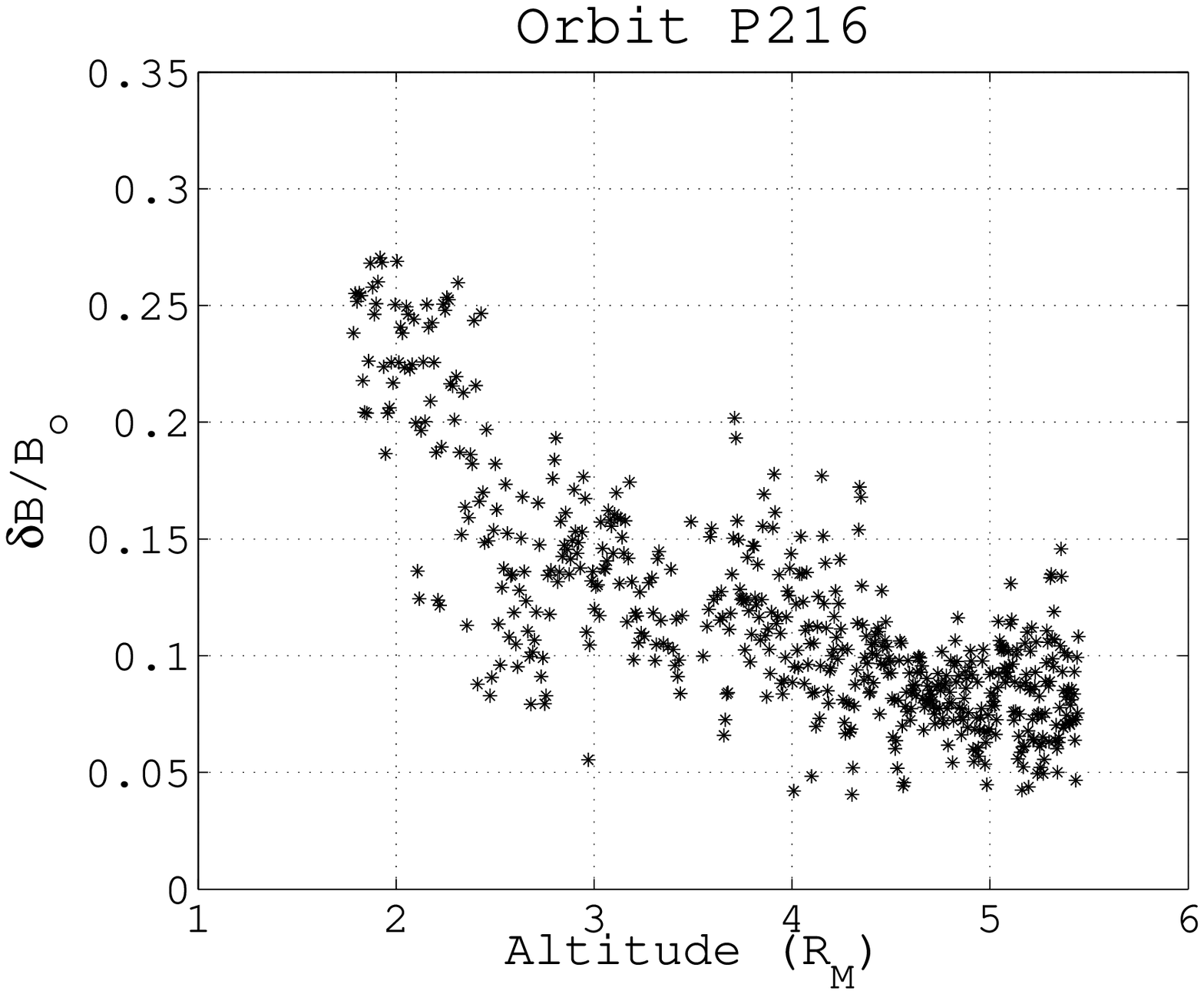}
\caption{ The relative wave amplitudes decrease with radial distance 
from the planet. Orbit P216, (April 3, 1998). An exponential fit $\,y = A\, exp (-\frac{r}{a})$ yields 
a scale height $a = (4.15 \pm 1.5) \, R_{M}$.}
 \label{amplitudes}
\end{center}
\end{figure}

We also see that the wave amplitudes decrease with radial distance
 from the planet, supporting the idea that Mars is the source of these waves. Figure \ref{amplitudes} illustrates what 
is observed in orbit P216. A similar behavior is observed for other orbits.

\subsection{ IMF dependence: Connection to the Martian bow shock}

\begin{table*}[ht]

\centering
\begin{tabular}{c c c c c c c c c}
\hline\hline
{Orbit}  & {D. Pol} & {Ell} & {Coh} & {$\lambda_{2}/\lambda_{3}$} & {{$\theta_{kB}$ (\textdegree)}} & {\textbar B\textbar} & {{$\alpha_{V,B}$ (\textdegree)}} &  {Connection}\\ [0.5ex]
\hline
{204} & {94.7} & {-0.72}& {0.90}& {31.4} & {8.9}& {4.86}& {51.9}& {Yes}\\

{207} & {95.4} & {-0.72}& {0.90}& {32.8} & {14.2}& {5.94}& {59.6} & {No}\\ 
{208} & {89.6} & {-0.67}& {0.80}& {17.4} & {10.0}& {5.04}& {54.1} & {Yes}\\ 
{209} & {88.6} & {-0.67}& {0.78}& {13.9} & {19.3}& {9.04}& {49.8} & {Yes}\\
{211} & {91.7} & {-0.68}& {0.83}& {15.3} & {15.6}& {4.65}& {56.6} & {Both}\\ 
{215} & {85.4} & {-0.65}& {0.73}& {18.1} & {38.4}& {13.5}& {40.2} & {No}\\  
{216} & {94.5} & {-0.69}& {0.89}& {20.2} & {17.6}& {7.31}& {33.6} & {Both}\\ 
{217} & {93.3} & {-0.70}& {0.84}& {19.6} & {24.3}& {4.62}& {31.5} & {No}\\ 
{239} & {93.3} & {-0.70}& {0.87}& {37.6} & {6.14}& {4.10}& {65.2} & {No} \\ 
{241} & {91.5} & {-0.68}& {0.83}& {32.1} & {17.8}& {7.20}& {76.9} & {No}\\ 
{242} & {91.5} & {-0.69}& {0.81}& {19.7} & {20.6}& {15.0}& {71.7} & {No}\\ 
{249} & {87.1} & {-0.63}& {0.76}& {10.0} & {29.6}& {4.87}& {32.6} & {Both}\\ 
{257} & {86.4} & {-0.64}& {0.75}& {16.1} & {15.2}& {4.44}& {61.7}& {Yes}\\ 
{258} & {90.8} & {-0.68}& {0.82}& {30.4} & {21.9}& {4.63}& {54.1}& {No}\\ 
{260} & {90.6} & {-0.67}& {0.82}& {33.3} & {14.1}& {5.27}& {42.8}& {Yes}\\ 
{267} & {91.7} & {-0.70}& {0.83}& {42.6} & {19.6}& {5.02}& {68.4}&  {No}\\ 
{268} & {89.4} & {-0.67}& {0.80}& {23.1} & {12.7}& {5.06}& {24.4}&  {Yes}\\ 

\hline
\end{tabular}
\label{table_wavesprop}
\caption{List of some selected orbits. Orbit number and wave properties. From left to right: degree of polarization, ellipticity, coherence,  $\lambda_{2}/\lambda_{3}$, $\theta_{kB}$, 
 \textbar B\textbar, Cone Angle $\alpha_{V,B}$, Connection to the Martian Bow Shock (Yes, No, Both cases).}
\end{table*}

The properties of PCW's are the same
inside and outside the nominal Martian foreshock. The results are summarized in Table 1 which shows the
average properties of PCWs for selected orbits that were chosen based on their
particularly high values of degree of polarization and coherence.
For example, orbits P204 and P207 have similar wave properties: they both are characterized 
by a high degree of polarization ($\geq$ 90\%), a large coherence (0.9) and a left-hand ellipticity (-0.72). $\theta_{kB}$
 is less than 15\textdegree  and 
their IMF cone angles differ in approximately 8\textdegree. Furthermore,  
in both cases the waves are observed in the same region (approximately between (2.09,-1.27,-2.22) $R_{M}$ and 
(1.7,-1.38,-4.21) $R_{M}$ in MSO coordinates).
However, whereas in P204 the magnetic field lines are connected to the bow shock, this does not occur in P207. The difference 
arises because in the first case, $\vec{B}_{o}$ is approximately (3,-1.5,-3) nT and in the second one (3,-5,0.7) nT. 
This strongly suggests that these are not
associated with Solar Wind backstreaming ions within the ion Martian
foreshock.

\subsection{Overall occurrence} \label{ovocc}

A study of the overall occurrence demanded the definition of a few criteria of the type of oscillation that would be considered to be a PCW.
These criteria are:

\begin{itemize}
\item Frequency close to the calculated proton cyclotron frequency ($\pm 1$nT).
\item MVA: events with $\lambda_{2}/\lambda_{3}\geq10 $.
 \item Events with coherence$\geq 0.7 \,$, ellipticity $\leq -0.5\,$, degree of polarization $\geq 70\%$.
\end{itemize}

As we explained in section (2), we study the temporal variation of the different waves properties breaking time series 
into 10-minute overlapping segments.

{ Figure \ref{amplaltid} shows the wave amplitude versus planetocentric distance for all SPO orbits.  
The amplitude is estimated from the power spectral density in an interval centered at the proton
cyclotron frequency with a width of 0.015 Hz. Cyan 
crosses correspond to SPO1, while brown open circles indicate wave occurrence for SPO2. The black curve displays the average amplitudes 
for $0.5 \,R_{M}$ bins, showing a general decrease of amplitude with increasing distance. Assuming that the source 
for these waves are exospheric pick-up protons, the obtained result is to be expected for at least two reasons: the ion density decreases 
with distance, and for smaller  MGS altitudes, waves have more time to grow as they are convected by the Solar Wind.

\begin{figure*}[ht]
\begin{center}
\includegraphics[scale=.5]{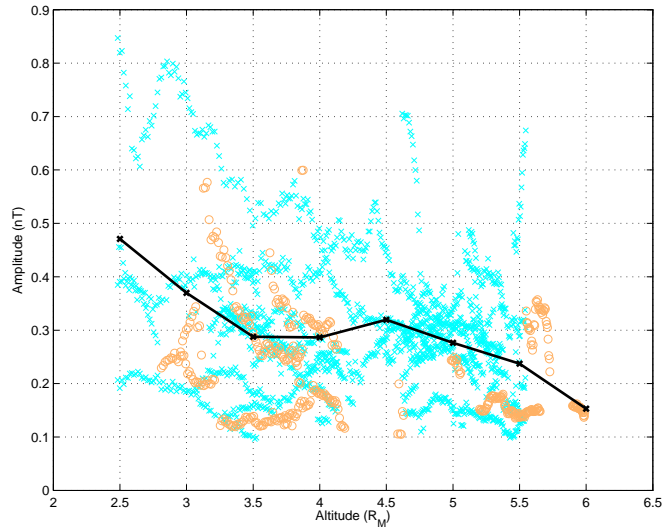}
\caption{ Amplitude of the PCW's as a function of the altitude for all SPO orbits. (Criteria stated at the beginning of section \ref{ovocc}).}
 \label{amplaltid}
\end{center}
\end{figure*}

{ Figure \ref{amplaphn} shows the amplitude of the waves (estimated in the same way as in 
figure \ref{amplaltid}) as a function of the observed IMF cone angle for all SPO orbits.  
The  cyan crosses correspond to SPO1 and the brown open circles to SPO2, while the black curve displays the 
average amplitudes for 10\textdegree$\,$ bins. 
This figure does not show any statistical shift between these two populations. 
The 10\textdegree $\,$ bin curve shows a slight shift toward small amplitudes as the IMF cone angle increases.
This trend is similar to the behavior
of the saturation amplitude of the instablity shown in \cite{cowee}.
However, \cite{cowee} suggested that the saturation level is likely not reached in the upstream region of Mars, 
which is not necessarily the case for the waves measured by MGS.
The observed differences could be due to the non homoscedasticity of the $\alpha_{V,B}$ distribution 
and the $\alpha_{V,B}$ estimation errors, which make direct comparison with their results not so straightforward.
}

\begin{figure*}[ht]
\begin{center}
\includegraphics[scale=.62]{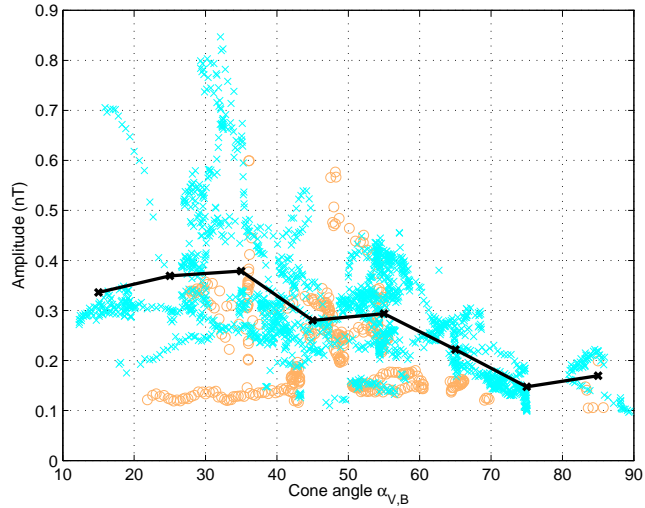}
\caption{ Amplitude of the PCW's as a function of the IMF cone angle for all SPO orbits. (Criteria stated at the beginning of section \ref{ovocc}).}
 \label{amplaphn}
\end{center}
\end{figure*}

 We also analyze the spatial distribution of the orbit segments
where waves were found,  making use of the MBE coordinate system. 
Figure \ref{mbeac12}.a shows the wave amplitudes (color coded) for 
the projection of the trajectory  on the  ($\vec{Y}_{MBE},\vec{Z}_{MBE}$) plane for all SPO1 and SPO2 orbits. Figure \ref{mbeac12}.b 
shows the  $\alpha_{V,B}$ value instead. 
In the absence of velocity vector measurements,  $\alpha_{V,B}$  is obtained assuming that $\vec{V}_{SW}$ is
parallel to the Sun-Mars line and therefore:  $\alpha_{V,B}=acos(\frac{B_{XMSO}}{B}).$
In both  figures,  closed and open circles correspond to SPO1 and SPO2 respectively.
The analyses do not show any clear difference in the spatial distribution of the waves between these two sub-phases:
 60\% of the waves occur above the $Z_{MBE} = 0$ plane and 40\% are below it, while the 
spacecraft is 48\%  and 52\% of the time in each hemisphere. As a result, the spatial distribution of PCWs does not 
seem to be affected by the Solar Wind's convective electric field.  
In fact, the spatial occurrence of the amplitude of the waves and the cone angle $\alpha_{V,B}$ associated with them 
are not affected by the electric field either. Moreover, these waves are observed even
 when $\vec{E}_{c}$ is relatively weak.

 \begin{figure}
\centering
  \begin{tabular}{cc}

\epsfig{file=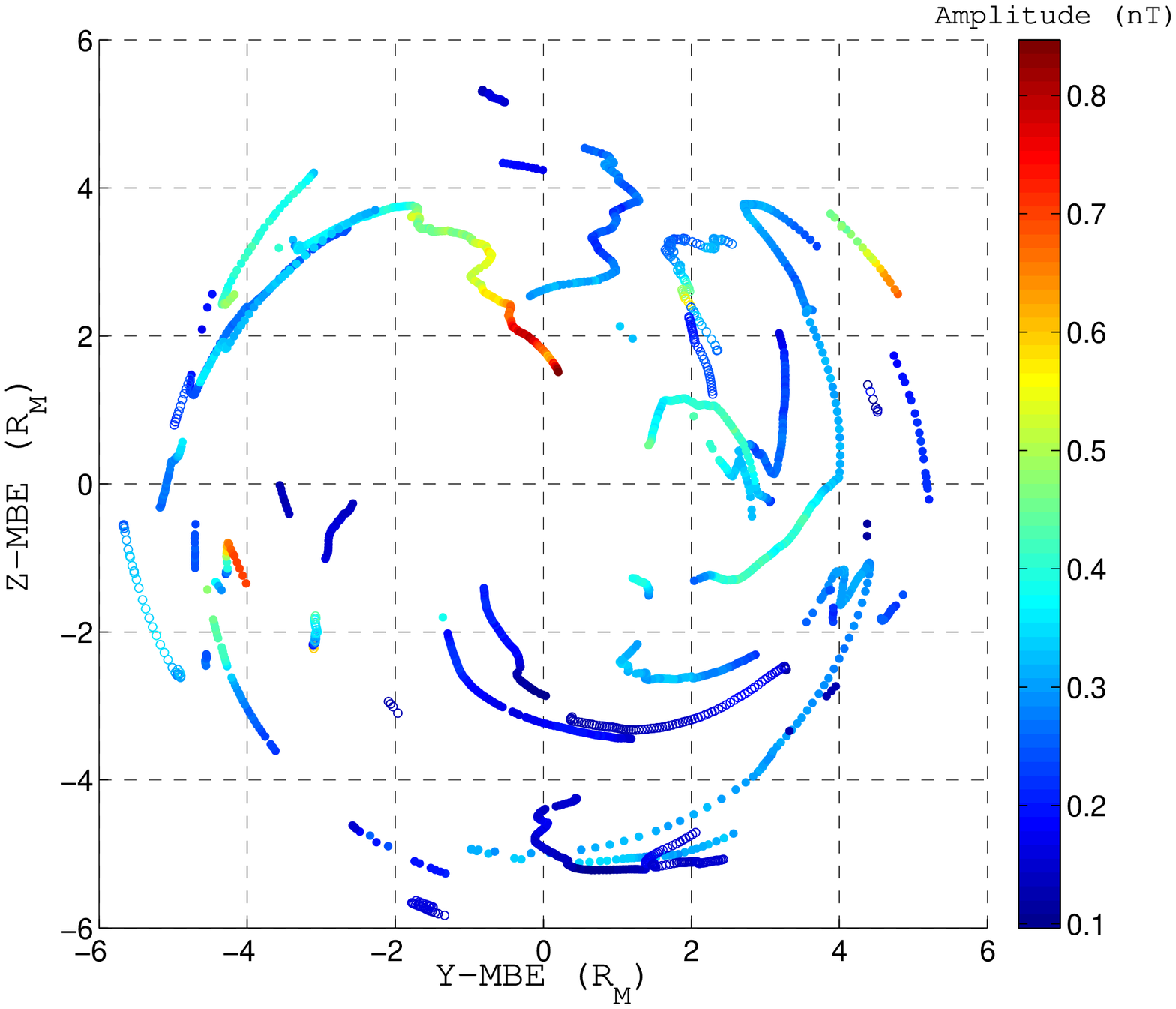,scale=0.48,clip=} \\
\epsfig{file=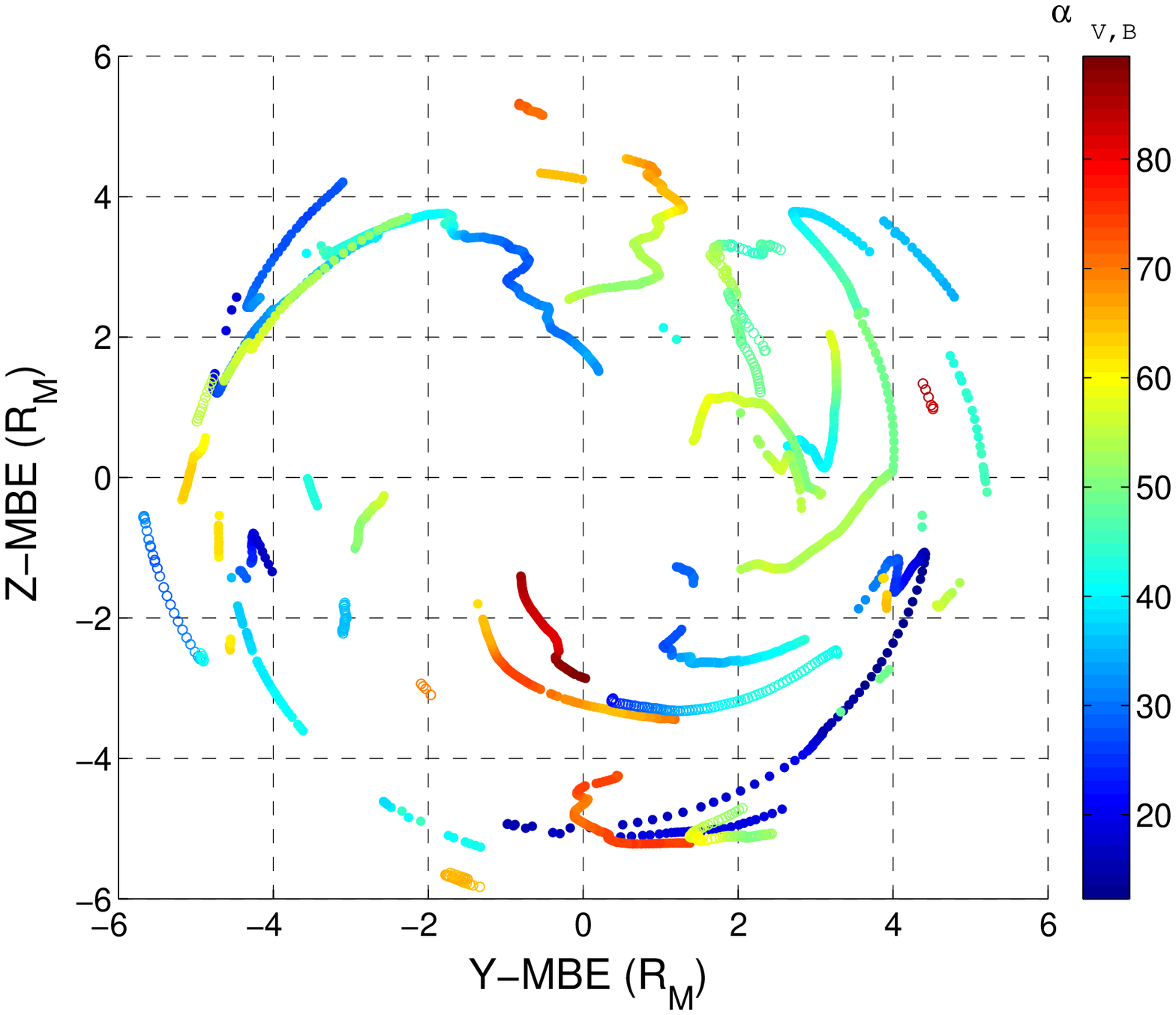,scale=0.50,clip=} &
\end{tabular}
\caption{Trajectory of the spacecraft SPO orbits with PCW's (MBE coordinate frame). Upper panel (a): Color coded by the amplitude of the waves.
Lower panel (b): Coded by color by the cone angle associated with the MAG measurements where it is observed PCW's. Closed and open circles
correspond to SPO1 and SPO2, respectively.}
\label{mbeac12}

\end{figure}

{ Another interesting aspect of this study is the clear difference in the occurrence rate of the waves between SPO1 and SPO2: 
whereas PCWs are present in 62\% of the SPO1 upstream observations, they only appear in 8\% of the time spent by MGS in the 
upstream region during SPO2.}
The waves during SPO1 last for longer and have a higher degree of polarization and coherence than those observed 
in SPO2.

{ This difference in the occurrence of PCW's is in principle not due to changes in the $\alpha_{V,B}$ as the distributions for 
both subphases peak at around 60\textdegree (Parker's spiral
angle at Mars 55\textdegree) with a standard deviation of 19\textdegree. }

{

}

\section{Discussion - Planetary Ion Pick-up and Waves at the Cyclotron Frequency}
\label{discusion}

One of the main results of the present study is
that the waves in the region upstream of the Martian bow shock
are observed, in the SC 
frame, at the local proton cyclotron frequency and are left-hand elliptically polarized.

As described in the introduction section, the Vlasov-Maxwell theory provides a mechanism capable of generating this type of 
waves as a result of the interaction between the magnetic field and 
the distribution function of \textquotedblleft exospheric \textquotedblright $\,$pick-up protons
  in the Solar Wind \citep{russell90}. { In particular, numerical studies \citep{gary93} 
  suggest two possible modes LH and RH whose predominance depends on the initial distribution function of the pick-up
  protons, which in turn, depends on the IMF cone angle.
  The dispersion obtained  in the values of the IMF cone angle (where $\vec{V}_{SW}$ was assumed to be along the Sun-Mars 
  direction) indicates that in principle both modes could be present in MAG observations.
  On the other hand, all wave observations in the SC frame displays a LH polarization. In the absence of ion measurements it is 
  impossible to obtain a value for the phase velocity of these waves. In particular, the angle between $\vec{k}$ and $\vec{V}_{SW}$ (that is necessary
  to compute the Doppler correction) at which
 the polarization changes remains unknown.
  However, following theoretical results \citep{cowee} one could assume that the phase speed is of the order of the 10\% of the ${V}_{SW}$, 
  in which case the cutoff angle would be around 95\textdegree. 
  In that case, most of the waves observed would correspond to the RH mode.
  LH waves have been observed at comet Grigg-Skjellerup \citep{neubauer93} for IMF cone angles near 90\textdegree [\cite{cao98} and references therein].}

MVA analysis shows that the observed PCW's are planar and propagate almost parallel (but not exactly) to $\vec{B}$. 
This could be the result of nonlinear effects
not addressed in this paper. This non zero $\theta_{kB}$ is responsible for a small compressibility
 (typically, $\delta B_{par}/ B \simeq 0.25$).

The compressibility of these waves was studied from the cross-correlation between the fluctuating
electron flux (with the electron fluxes taken as a proxy of the electron density) and the parallel magnetic field component. 
The maximum values for the cross-correlation coefficient between both time series are associated with  
phase shifts between 0 and 2s.
The quality of this result is limited by the following two factors: the influence of the magnetic fields generated 
by the spacecraft and the time resolution of the ER instrument(2s).
These two sources of error impose a tradeoff in our aim to obtain the best possible cross-correlation between the two time series. 
With this in mind we calculated the cross-correlation for several intervals with mean magnetic fields between 4 and 8 nT.  
For all the cases considered, the results indicate a maximum in the correlation coefficient 
located within 2s from zero time lag. This means that there is a systematic correlation between the two data sets within the errors caused by the 
spacecraft fields and the ER time resolution.  But also, these results indicate that anticorrelation is not possible (errors are not
 as high as half a proton cyclotron period). 

The analysis of MAG measurements also shows that the frequency, polarization and coherence
of these waves do not seem to depend on the direction of the IMF (see Table 1). It is then not surprising that
the observed waves are not associated with the foreshock. In addition, the amplitude of the observed waves
shows a decrease as the IMF cone angle increases.

The lack of influence of $\vec{E}_{c}$ on the spatial distribution of PCWs { as well as their amplitudes} suggests that
the link between the spatial distribution of the { pick up} ions and the waves is not obvious \citep{cowee}. 
On the other hand, the decrease of the observed wave amplitude with radial distance
 from the planet, supports the idea that Mars is indeed the source of these waves.

We also analyzed the PCWs overall occurrence using very strict criteria. We found 
a very clear difference in the percentage and the properties of PCWs between SPO1 and SPO2. 
However, a statistical study of the IMF cone angle did not show any significant difference between both periods, 
suggesting that the pick-up geometry might have not changed significantly from SPO1 to SPO2.
The difference in the occurrence of the waves between SPO1 and SPO2 might be related to 
temporal or spatial changes in the properties of the Martian hydrogen exosphere.

\cite{wr2006} studied the wave occurrence in AB1, as seen from a reference frame that takes into account the convective 
electric field. They observed that the PCWs at large distances 
(85 events present in 9 orbits with $B >5.6$ nT) occur intermittently { and predominantly in the hemisphere where $\vec{E}_{c}$ points 
out from the planet (+E hemisphere)}. In order to explain this behavior the authors proposed a mechanism where an exospheric hydrogen atom is ionized 
 and then accelerated by the Solar Wind electric field. A charge exchange collision would then produce fast neutrals able to reach
distant regions where they are re-ionized and generate cyclotron waves downstream of the planet and on the + E hemisphere.

During SPO1 and SPO2 the wave spatial distribution for distances smaller than 6 $R_{M}$ does not seem to depend on the orientation of
the Solar Wind convective electric field (even after more restrictive criteria have been applied). 
Furthermore, waves are found even when $\vec{E}_{c}$ is relatively weak.
 We also observe PCWs far away
 into the -E hemisphere, although there is no known mechanism to move ions against the 
electric field. 
One explanation to this is that the wave distribution does not necessarily follow the one of pick-up ions. 
It is important to note that pick-up ions generate waves whose wavelengths are of planetary scale. Ion density is not 
likely to be homogeneous over such large spatial distances.
The difference between the results obtained in \cite{wr2006} and the ones presented here
could be due to the fact that MGS sampled different regions during 
AB1 and SPO. On the other hand it is also important to point out that the lack of correlation between the spatial distribution of the 
PCWs and the convective electric field has also been observed at Venus \citep{delva2011}.

\section{Conclusions}
\label{conclusiones}

In this study we analyzed upstream MGS MAG/ER measurements during the pre-mapping SPO phase.
Analyses of these measurements show waves whose frequencies in the reference 
system of the satellite are near the local cyclotron frequency of protons. 
These ultra low frequency plasma waves are also characterized (in the
spacecraft frame) by a left-hand elliptical polarization and by an amplitude that decreases {with the increase of the IMF cone angle and the}
radial distance from the planet. 
They propagate almost parallel to the background magnetic field and show
a small degree of compressibility. The observed waves are not associated to the foreshock and their properties do not depend
on the angle $\alpha_{vB}$ between the solar wind velocity and the magnetic field direction.

The most plausible explanation for the existence of these waves is that they are generated from the pick-up of newborn protons from the
Martian exosphere whose distribution function kinetically interacts with the Solar Wind plasma and the interplanetary magnetic field \citep{russell90}.

{ From a theoretical perspective, there are two possible modes that can be generated in the plasma frame which
depends on the pick-up proton distribution function.
For relative orientations between $\vec{V}_{SW}$ and $\vec{B}$ ranging from parallel to almost perpendicular, the RH 
mode will predominate, while for almost perpendicular orientations the LH mode will dominate.
The reason why we always observe left-hand polarized waves in the SC frame is because of their corresponding Doppler corrections.}

The spatial distribution of these waves does not seem to depend on the orientation of
the Solar Wind convective electric field and they are also observed even
 when it is relatively weak. PCWs are also found far
 into the negative convective electric field region. 
This supports the idea from recent { simulation} results \citep{cowee} that, even though the origin of these waves are
the pick-up ions, there is no clear relation between their corresponding spatial distributions.

We also found a clear difference in the PCWs occurrence between SPO1 and SPO2. 
Moreover, it is observed that the waves observed during SPO1 last longer, and have higher degree
of polarization and coherence than those observed in SPO2.
Since there is no significant difference in the IMF cone angle distribution between both periods, 
these differences might be due to changes in the properties of the exosphere and/or ionization rates.

A more accurate determination of the wave modes present at Mars will require information about the
pick up ion distribution function as well as that of the Solar Wind. We hope
MAVEN spacecraft (Mars Atmosphere and Volatile Evolution Mission) will provide such measurements contributing 
to a better understanding of the microscopic processes arising from the interaction between the Solar Wind and Mars.




\section*{Acknowledgments}

The authors also wish to thank Misa Cowee for helpful and interesting suggestions and comments. 
This work was done in conjunction with the International Space Science Institute (ISSI) Working
Group on Induced Magnetospheres. N.R. is supported by a National Science and Technology Research Council (CONICET) Phd grant.


\begin{thebibliography}{00}


\bibitem [Acu\~na et al.(1992)]{acuna92}  {Acu\~na, M., et al., 1992. The Mars Observer magnetic fields investigations. \textit{J. Geophys. Res.} 97, E5, 7799-7814.}
%
\bibitem [Acu\~na et al.(1998)] {acuna98} {Acu\~na, M., et al., 1998. Magnetic field and plasma observations at Mars: initial results of the Mars Global Surveyor Mission. Science 279, 1676-1680.}
%
\bibitem[Acu\~na et al.(2001)]{acuna2001} {Acu\~na, M., et al., 2001. Magnetic field of Mars: summary of results from the aerobraking and mapping orbits. \textit{J. Geophys. Res.} 106, E10, 23403-23417.}


\bibitem [Albee et al.(2001)] {albee2001} {Albee, A., et al., 2001. Overview of the Mars Global Surveyor mission. \textit{J. Geophys. Res.} 106, E10, 23291-23316.}

%
\bibitem [Bertucci et al.(2005)]{bertucci2005} {Bertucci, C., et al., 2005. Interaction of the Solar Wind with Mars from Mars Global Surveyor MAG/ER observations. J. Atmospheric and Solar-Terrestrial Physics 67, 1797-1808.}
%
\bibitem [Brain et al.(2002)]{brain2002} {Brain, A., et al., 2002. Observations of low-frecuency electromagnetic plasma waves upstream from the Martian shock. J. Geophys. Res. 107, A6, 1076.}
%
\bibitem  [Brinca and Tsurutani(1989)]{tsubrinc89} {Brinca, A., Tsurutani, B.T., 1989. Influence of multiple ion species on low-frequency electromagnetic wave instabilities, \textit{J. Geophys. Res.}, 94, 13565-13569.}


\bibitem [Brinca(1991)] {brinca1991} {Brinca, A., 1991. Cometary linear instabilities: from profusion to perspertive in Cometary Plasma Processes. Geophys. Monograph 61, 211-221.}

\bibitem [Cao et al.(1998)] {cao98}{Cao, J.B., Mazelle, C., Belmont, G., Reme, H.: J. Geophys. Res. 103,2055 (1998)}

\bibitem [Chaufray et al.(2008)] {Chaufray} {Chaufray, J.Y., Bertaux, J.L., Leblanc, F., Quémerais, E., 2008. Observation of the hydrogen corona with SPICAM on Mars Express. Icarus, 195, Issue 2, 598-613.}

%
\bibitem [Convery and Gary(1997)]{convgary97} {Convery, P., and Gary, P., 1997. Electromagnetic proton cyclotron ring instability: Threshold and saturation. J. Geophys. Res. 102, A2, 2351-2358.}

\bibitem[Cowee et al.(2012)]{cowee}{Cowee, M. M., S. P. Gary, and H. Y. Wei (2012), Pickup ions and ion cyclotron wave amplitudes upstream of Mars: First results from the 1D hybrid simulation, Geophys. Res. Lett., 39, L08104, doi:10.1029/2012GL051313.}


\bibitem [Delva et al.(2009)]{delva09} {Delva, M., et al., 2009. Hydrogen in the extended Venus exosphere. \textit{J. Geophys. Res.} 36, L01203.}
%

\bibitem[Delva et al.(2011)]{delva2011}{Delva, M., Mazelle, C., Bertucci, C., Volwerk,M.,  V\"{o}r\"{o}s, Z., Zhang, T.L., 2011. Proton cyclotron wave generation mechanisms upstream of Venus. J. Geophys. Res. 116, A02318, doi:10.1029/2010JA015826.}

\bibitem[Fowler et al.(1967)]{fowler}{Fowler R. A., Kotick B.J., Elliot D., 1967. Polarization Analysis of Natural and Artificially Induced GeomagneticMicropulsations. \textit{J. Geophys. Res.} 72, No. 11.}

\bibitem [Gary(1991)]{gary91} {Gary, S.P., 1991. Electromagnetic ion/ion instabilities and their consequences in space plasmas: A review, Space Sci. Rev., 56, 373–415.}

\bibitem [Gary(1993)]{gary93} {Gary, S.P., 1993. Theory of Space Plasma Microinstabilities. In: Cambridge Atmospheric and Space Series. Cambridge University Press, Cambridge.}

\bibitem [Gary and Madland(1988)]{garymad} {Gary, S.P., and Madland, C.,  1988. Electromagnetic ion instabilities in a cometary environment. \textit{J. Geophys. Res.} 93, A1, 235-241.}

\bibitem [Gary et al.(1989)]{gary89} {Gary, S.P., Akimoto, K., Winske D.,  1989. Computer simulations of cometary ion/ion instabilities and wave growth. \textit{J. Geophys. Res.} 94, A4, 3513-3525.}

\bibitem [Khrabrov and Sonnerup(1998)]{mvab} {Khrabrov, A., and Sonnerup, B., 1998. Error estimates for minimum variance analysis. J. Geophys. Res. 103, A4, 6641-6651.}

\bibitem [Knetter et al.(2004)] {knetter04} {Knetter, T., et al., 2004. Four-point discontinuity observations using Cluster magnetic field data: A statical survey. J. Geophys. Res. 109, A06102.}


\bibitem [Mazelle and Neubauer(1993)]{mazelleneubabuer93}{Mazelle, C., and Neubauer, F.M., 1993. Discrete wave packets at the proton cyclotron frequency at Comet P/Halley.Geophys. Res. Lett. 20, 153.}


\bibitem[Mazelle et al.(2004)]{mazelle2004}{Mazelle, C., et al., 2004. Bow shock and upstream phenomena at Mars, Space Science Review 111, 115-181.}

\bibitem[McPherron et al.(1972)]{mcpherron}{McPherron, R.L., Russell,C.T., Coleman, P.J., 1972. Fluctuating magnetic fields in the magnetosphere, II, ULF waves. Space Sci. Rev. 13, 411-454.}

\bibitem [Mitchell et al.(2001)] {er} {Mitchell, D., et al., 2001. Probing Mars' crustal magnetic field and ionosphere with the MGS Electron Reflectometer. J. Geophys. Res. 106, E10, 23419-23427.}

\bibitem[Modolo et al.(2005)]{modolo05}{Modolo, R., Chanteur, G. M., Dubinin, E., Matthews, A.P., 2005. Influence of the solar EUV flux on the Martian plasma environment. Annales Geophysicae 23, 433–444.} 

\bibitem[Neubauer et al.(2003)]{neubauer93}{Neubauer, F. M., K. .-H. Glassmeier, A. J. Coates, and A. D. Johnstone (1993), Low-Frequency Electromagnetic Plasma Waves at Comet P/Grigg-Skjellerup: Analysis and Interpretation, J. Geophys. Res., 98(A12), 20,937–20,953, doi:10.1029/93JA02532.}

\bibitem  [Rankin and Kurtz(1970)] {rk1970} {Rankin, D., and Kurtz, R., 1970. Statistical study of micropulsation polarizations. \textit{J. Geophys. Res.} 75, 5444-5458.}
%
\bibitem  [Russell et al.(1990)]{russell90} {Russell, C.T., et al., 1990. Upstream waves at Mars - PHOBOS observations. \textit{Geophysical Research Letters} 17, 6, 897-900.}




\bibitem  [Smith and Tsurutani(1976)]{smith76} {Smith, E.J., and Tsurutani, B.T., 1976. Magnetosheath Lion Roars, \textit{J. Geophys. Res.} 81, 2261-2266.}


\bibitem [Sonnerup and Scheible(1998)]{sonerup}{Sonnerup, B. U. O., Scheible, M., 1998. Minimum and Maximum Variance Analysis, in: Paschmann, G., Daly, P. W.
(Eds), Analysis Methods for Multi-Spacecraft Data. Int. Space Sci. Inst., Bern, Switzerland.}



\bibitem  [Tsurutani and Smith(1986)]{tsu1986} {Tsurutani, B.T., and Smith, E.J., 1986. Strong hydromagnetic turbulence associated with comet Giacobini-Zinner, Geophys. Res. Lett., 13, 259-262.}

\bibitem  [Tsurutani et al.(1987)]{tsu1987} {Tsurutani, B.T., et al., 1987. Steepened magnetosonic waves at comet Giacobini-Zinner, \textit{J. Geophys. Res.} 92, 11074-11082.}

\bibitem  [Tsurutani et al.(1989)]{tsu1989} {Tsurutani, B.T., et al., 1989. Magnetic Pulses with Durations near the Local Proton Cyclotron Period: Comet Giacobini-Zinner, \textit{J. Geophys. Res.}, 94, 29-35.}

\bibitem  [Tsurutani(1991)]{tsu1991}{Tsurutani, B. T., 1991. Comets: A laboratory for plasma waves and instabilities, 
in Cometary Plasma Processes, Geophys. Monogr. Ser., 61, 189–209, AGU, Washington, D. C.}


\bibitem[Tsurutani et al.(2009)]{tsu2009}{Tsurutani, B. T., O. P. Verkhoglyadova, G. S. Lakhina, and S. Yagitani (2009), Properties of dayside outer zone chorus
during HILDCAA events: Loss of energetic electrons, J. Geophys. Res., 114, A03207, doi:10.1029/2008JA013353.}

\bibitem [Vignes et al.(2000)]{vignes}{Vignes, D., et al., 2000. The Solar Wind interaction with Mars: locations and shapes of the Bow Shock and the Magnetic Pile-up Boundary from the observations of the MAG/ER 
experiment onboard Mars Global Surveyor. \textit{Geophysical Research Letters} 27, 1, 49-52.}

\bibitem [Wu and Davidson(1972)]{wu1972}{Wu, C. S., and R. C. Davidson (1972), Electromagnetic Instabilities Produced by Neutral-Particle Ionization in Interplanetary Space, J. Geophys. Res., 77(28), 5399–5406, doi:10.1029/JA077i028p05399.}

\bibitem [Wu and Hartle(1974)]{hartle74}{Wu, C.S., and Hartle, R.E. 1974. Further remarks on plasma instabilities produced by ions born in the solar wind. \textit{J. Geophys. Res.} 79, 283-285.}

\bibitem [Wei and Russell(2006)]{wr2006}{Wei, H., and Russell, C.T., 2006. Proton cyclotron waves at Mars: Exosphere structure and evidence for a fast neutral disk. \textit{J. Geophys. Res.} 33, L23103.}



 \end{thebibliography}
\end{document}